\documentclass[aps,prl,floatfix,twocolumn,superscriptaddress,showpacs,reprint]{revtex4}

\usepackage{graphicx}
\usepackage{mathrsfs}
\usepackage{amsmath}
\usepackage{amsfonts}
\usepackage{subfigure}
\usepackage{bm}
\usepackage{verbatim}
\usepackage{color}
\usepackage{xcolor}
\usepackage[colorlinks=true, letterpaper=true, pdfstartview=FitV, linkcolor=blue, citecolor=blue, urlcolor=blue]{hyperref}
\usepackage{siunitx}

\hypersetup{%
pdftitle={Quantum nonlinear Hall effect induced by Berry curvature dipole in time-reversal invariant materials},%
pdfauthor={Inti Sodemann and Liang Fu},%
pdfpagemode={UseNone},%
pdfstartview={FitH},%
breaklinks=true}

\newcommand{\be}{\begin{equation}}
\newcommand{\ee}{\end{equation}}
\newcommand{\ba}{\begin{equation}\begin{split}}
\newcommand{\ea}{\end{split}\end{equation}}

\begin{document}
\title{Quantum nonlinear Hall effect induced by Berry curvature dipole in time-reversal invariant materials}
\author{Inti Sodemann}
\affiliation{Department of Physics, Massachusetts Institute of Technology, Cambridge, Massachusetts 02139}
\author{Liang Fu}
\affiliation{Department of Physics, Massachusetts Institute of Technology, Cambridge, Massachusetts 02139}

\begin{abstract}
It is well-known that a non-vanishing Hall conductivity requires time-reversal symmetry breaking. However, in this work, we demonstrate that a Hall-like transverse current can occur in second-order response to an external electric field in a wide class of time-reversal invariant and inversion breaking materials,  at both zero and twice the optical frequency. This nonlinear Hall effect has a quantum origin arising from the dipole moment of the Berry curvature in momentum space, which generates a net anomalous velocity when the system is in a current-carrying state. We show that the nonlinear Hall coefficient is a rank-two pseudo-tensor, whose form is determined by point group symmetry. We discus optimal conditions  to observe this effect and propose candidate two- and three-dimensional materials, including topological crystalline insulators,  transition metal dichalcogenides and Weyl semimetals.
\end{abstract}
\maketitle

\noindent
{\color{blue}{\em Introduction}}---The Hall conductivity of an electron system whose Hamiltonian is invariant under the time reversal symmetry is forced to vanish.  Crystals with sufficiently low symmetry can have resistivity tensors which are anisotropic, but, Onsager's reciprocity relations~\cite{LandauLifshitz} force the conductivity to be a symmetric tensor in the presence of time reversal symmetry, implying that, when the electric field is along its principal axes, the current and the electric field are collinear, at least to the first order in electric fields. However, this constraint is only in linear response theory and does not necessarily enforce the full current to flow collinearly with the local electric field. 

In this paper we study a special type of such non-linear Hall-like currents. We will demonstrate that metals without inversion symmetry can have a non-linear Hall-like current that arises from the Berry curvature in momentum space. The conventional Hall conductivity can be viewed as the zero order moment of the Berry curvature over occupied states, namely, as an integral of the Berry curvature within the metal's Fermi surface in momentum space. The effect we discuss here is determined by a pseudo-tensorial quantity that measures a first order moment of the Berry curvature over the occupied states, namely its dipolar distribution, and hence we call it the Berry curvature dipole. This nonlinear Hall effect has a {\it quantum} origin arising from the anomalous velocity of Bloch electrons generated by the Berry curvature~\cite{Sundaram1999}, but it is not expected to be quantized.

In a time reversal invariant system, the Berry curvature is odd in momentum space, $\Omega_a(k)=-\Omega_a(-k)$, and hence its integral weighed by the equilibrium Fermi distribution is forced to vanish, because states at $k$ and $-k$ form Kramers pairs and are equally occupied. 
However, the second order response is determined by the integral of the Berry curvature evaluated in the {\it non-equilibrium} distribution of electrons computed to first order in the electric field. Since the non-equilibrium current-carrying distribution is not symmetric under $k \rightarrow -k$, the integral of the Berry curvature weighed by it can be finite, leading to a net anomalous velocity and hence a transverse current.  

Our study builds upon a seminal work by Moore and Orenstein~\cite{Moore2010}, which predicted a DC photocurrent in quantum wells without inversion symmetry due to the anomalous velocity associated with the Berry phase. The quantum nonlinear Hall effect presented here can be regarded as a generalization of this effect. We  predict that an oscillating electric field can generate a transverse current at {\it both zero and twice the frequency} in two- and three-dimensional materials with a large class of crystal point group symmetries. In particular, 
the second harmonic generation is  a distinctive signature that may facilitate the experimental detection of the quantum nonlinear Hall effect. 
Additionally, the effect does remain finite in the dc limit of the applied electric field.

{\color{blue}{\em General theory}}---The electric current density is given by the integral of the physical velocity of the electrons, $v_a$, weighed by their occupation function, $f(k)$:
\be
j_a=-e \int_k f(k) \ v_a.
\ee 
For simplicity we imagine a single band system but allow it to be two- or three-dimensional: $\int_k\equiv \int d^d k/(2\pi)^d$. The velocity contains two contributions, namely, the group velocity of the electron wave and the anomalous velocity arising from the Berry curvature~\cite{Sundaram1999} ($\hbar=1$):
\be\label{vel}
v_a=\partial_a \epsilon(k)+ \varepsilon_{abc} \Omega_b  \dot{k}_c,
\ee 
where $\epsilon$ and $\Omega_b$ are the energy dispersion and the Berry curvature of the electrons in question:
\begin{equation}\label{Berry}
\Omega_a \equiv\varepsilon_{abc} \partial_b A_c,\ A_c \equiv -i\langle u_k|\partial_c|u_k\rangle.
\end{equation}
Within the Boltzmann picture of transport, the canonical momentum of electrons changes in time in response to the external electromagnetic fields. In the absence of external magnetic fields the change of momentum is:
\be
\dot{k}_c=-e E_c(t).
\ee
where $E_c(t) =\Re\{\mathcal{E}_c e^{i \omega t}\}$, with $\mathcal{E}_c \in \mathbb{C}$, is the driving electric field which oscillates harmonically in time but is uniform in space. In the relaxation time approximation the Boltzmann equation for the distribution of electrons is~\cite{Mahan}:
\be
-e\tau E_a \partial_a f+\tau \partial_t f=f_0-f,
\ee
%\be
%-\frac{e}{\hbar} E_a \partial_a f+\partial_t f=-\frac{f-f_0}{\tau},
%\ee
where $f_0$ is the equilibrium distribution in the absence of external fields. We are interested in computing the response to second order in the electric field, hence we expand the distribution up to second order: $f=\Re\{f_0+f_1+f_2\}$, where the term $f_n$ is understood to vanish as $\mathcal{E}^n$. One finds a recursive structure:
\begin{equation}\label{xx}
\begin{split}
f_1 & =f_1^\omega e^{ i \omega t},\ f_1^\omega =  \frac{e\tau \mathcal{E}_a \partial_a f_0}{1+i\omega\tau},\\
f_2 & =f_2^0+f_2^{2\omega} e^{2 i \omega t}, \ f^0_2 = \frac{(e\tau)^2 \mathcal{E}^*_a \mathcal{E}_b \partial_{ab} f_0}{2(1+i\omega\tau)},\\
f^{2\omega}_2 & =   \frac{(e\tau)^2 \mathcal{E}_a \mathcal{E}_b \partial_{ab} f_0}{2(1+i\omega\tau)(1+2 i\omega\tau)}.
\end{split}
\end{equation}
%\begin{equation}\label{xx}
%\begin{split}
%f_1 & =f_1^\omega e^{ i \omega t},\ f_1^\omega =  \frac{(e\tau/\hbar) \mathcal{E}_a \partial_a f_0}{1+i\omega\tau},\\
%f_2 & =f_2^0+f_2^{2\omega} e^{2 i \omega t}, \ f^0_2 = \frac{(e\tau/\hbar)^2 \mathcal{E}^*_a \mathcal{E}_b \partial_{ab} f_0}{2(1+i\omega\tau)},\\
%f^{2\omega}_2 & =   \frac{(e\tau/\hbar)^2 \mathcal{E}_a \mathcal{E}_b \partial_{ab} f_0}{2(1+i\omega\tau)(1+2 i\omega\tau)}.
%\end{split}
%\end{equation}
By writing the current as $j_a=\Re\{j^0_a+j^{2\omega}_a e^{2i\omega t}\}$, one obtains:
\begin{equation}\label{ja0}
\begin{split}
j_a^0&=\frac{e^2}{2} \int_k  \varepsilon_{abc}\Omega_b  \mathcal{E}_c^* f^{\omega}_1-e \int_k f_2^0 \partial_a    \epsilon (k)  
,\\
j_a^{2\omega}&=\frac{e^2}{2} \int_k  \varepsilon_{abc}\Omega_b\mathcal{E}_c f^{\omega}_1-e \int_k f_2^{2\omega} \partial_a  \epsilon (k).
\end{split}
\end{equation}

The term $j_a^0$ describes a rectified current while the term $j_a^{2\omega}$ describes a second harmonic generated by the non-linearity. The second terms that appear in both expressions above are completely semiclassical and do not require the presence of Berry curvature. However, within the approximation of a constant scattering time, one finds that these non-linear terms are proportional to the integral of a three-index tensor, $\partial_a  \epsilon (k) \partial_{bc} f_0(k)$, which is odd under time reversal and hence they are forced to vanish. Therefore the only surviving terms are those associated with the Berry curvature.  

By writing $j^0_a=\chi_{abc} \mathcal{E}_b \mathcal{E}^*_c$, $j^{2\omega}_a=\chi_{abc} \mathcal{E}_b \mathcal{E}_c$, one has that:
\be\label{chi}
\chi_{abc}=\varepsilon_{adc}  \frac{e^3\tau}{2(1+i \omega \tau)} \int_k   (\partial_b f_0) \Omega_d.
\ee
%Although the frequency dependence of the rectified current arising from the Berry curvature is the same as that arising from conventional second order term, the frequency dependence of the second Harmonic of the Berry curvature contribution  is clearly different from the second Harmonic of the conventional term. This could be employed to separate them experimentally. Additionally, the rectified Berry curvature term depends linearly on the scattering time $\tau$, while the rectified conventional term is quadratic in $\tau$. Moreover, as we will illustrate in the case of the surface of SnTe topological crystalline insulator (TCI), for certain crystal symmetries the convetional term vanishes and only the Berry curvature contribution survives. 
%In this work we will focus on the first term that requires the existence of a finite Berry curvature. The two terms will generally be present in materials, but as we will illustrate in an specific example in Sec. xxx, for certain crystal symmetries the second term will vanish. Even if the two terms are present they can be separated by their distinct dependences. In the dc limit $\omega\rightarrow 0$ the first term produces current that is always normal to the applied electric field: $j_a E_a=0$.
%Additionally, the first term depends linearly on the scattering time $\tau$, whereas the second is quadratic.
The presence of the factor $\partial_b f_0$ will gurantee that only states close to the Fermi surface will contribute to the integral in the low temperature limit, so that this response is a Fermi liquid property~\cite{Haldane2004}. Equation~\eqref{chi} can be rewritten as follows:
\be\label{chiabc}
\chi_{abc}=-\varepsilon_{adc} \frac{e^3 \tau}{2(1+i \omega \tau)} \int_k   f_0 \ (\partial_b \Omega_d).
\ee

This expression \eqref{chiabc} for the nonlinear conductivity tensor, $\chi_{abc}$, is the first main result of this work. It shows that $\chi_{abc}$ is proportional to the {\it dipole moment} of the Berry curvature over the occupied states, defined as: 
\be\label{Dab}
D_{ab}=\int_k   f_0 \ (\partial_a \Omega_b).
\ee
It is interesting to note that this tensor is dimensionless in three dimensions. 
At frequencies above the width of the Drude peak $\omega \tau \gg 1$ and below the interband transition threshold, 
the prefactor in $\chi_{abc}$ becomes independent of the scattering time, so that $\chi_{abc}$ is a direct measure of quantum geometry of the Bloch states in momentum space.  
 In the dc limit or for linearly polarized electric fields, the Berry curvature dipole term always produces a current that is orthogonal to the electric field $j_a E_a=0$~\footnote{Note that in the pure dc case, writing $E_c(t) =\mathcal{E}_c$, with $\mathcal{E}_c \in \mathbb{R}$, the current is
$j_a=2 \lim_{\omega\rightarrow 0}j_a^0$.}.

To close this section, we wish to remark that there exist additional second order corrections to the current that arise from modifications to Eq.~\eqref{vel} that are intrinsic to the band structure, containing no powers of the scattering time $\tau$~\cite{Gao2014}, however, these contributions vanish for time reversal invariant systems. Other type of rectifications might arise in systems with an inversion asymmetric scattering rate, namely one for which the scattering from $k$ to $k'$ has a different rate than that from  $-k$ to $-k'$, which produces a kind of ratchet effect~\cite{Olbrich2014}. These semiclassical Berry-phase independent contributions are distinguished from the quantum nonlinear Hall effect discussed in this work because they are expected to scale as $\tau^2$.

%Also, although we have restricted our discussion to a single-band systems for simplicity, we believe that extensions of our work to multiband systems is an interesting avenue for future of research.

%\begin{figure}[!]
%\includegraphics[scale=0.35]{Magnetic_BISFET-2.pdf}
%\caption{\label{Schematic}(color online). 
%Charge transport through two (Left and Right) metallic multilayer stacks containing both
%}
%\end{figure}

\noindent
{\color{blue}{\em Berry curvature dipole in three dimensions}}---%We define the Berry curvature dipole moment tensor as:
Let us explore the constraints imposed by crystal point symmetries on the Berry curvature dipole moment tensor $D_{ab}$. A point symmetry is described by an orthogonal matrix $S$. Because the Berry curvature is a pseudovector, the Berry curvature dipole transforms as a pseudotensor. Hence, crystal symmetries impose constraints of the form: 
\be\label{pseudoT}
D = \det(S) S D S^T.
\ee
To determine which components of this tensor are non-zero it is convenient to decompose it into symmetric and antisymmetric parts: $D^\pm=(D\pm D^T)/2$, which transform independently under symmetry operations. The antisymmetric part of a pseudotensor transforms as a {\it vector}, as can be verfied from Eq.~\eqref{pseudoT}. The components of this vector can be taken to be $d_{a}\equiv \epsilon_{abc} D^-_{bc}/2$. Therefore for it to be non-zero the crystal must have a polar axis. From the 32 crystallographic point groups, 10 allow for a polar axis, namely $\{C_n,C_{nv}\}$ with $n={1,2,3,4,6}$. The vector $d_{a}$ will be oriented along such axis. The contribution to the current from this antisymmetric part can be written in vector notation as:
\begin{eqnarray}\label{vecj}
\vec{j}^0&=&\frac{e^3\tau}{2 (1+i\omega\tau)} \vec{\mathcal{E}}^*\times(\vec{d}\times\vec{\mathcal{E}}), \nonumber \\
\vec{j}^{2\omega}&=&\frac{e^3\tau}{2 (1+i\omega\tau)} \vec{\mathcal{E}}\times(\vec{d}\times\vec{\mathcal{E}}). 
\end{eqnarray}

Let us now determine which crystals allow for a non-zero symmetric part $D^{+}$. We require the crystal to be inversion asymmetric for otherwise the Berry curvature would be identically zero due to the time reversal symmetry. Any real symmetric matrix can be diagonalized and has a real spectrum. Let us denote its eigenvalues and eigenvectors by $\delta_i$, ${\bf e}_i$ respectively: $
D^+=\sum_{i=1}^3 \delta_i {\bf e}_i {\bf e}_i^T$. All inversion asymmetric crystals without left-handed symmetries allow for $D^+$ to be non-zero, but might generally impose constraints on its eigenvectors to lie along the principal symmetry axis and some of its eigenvalues to be degenerate, much in the same way they constrain an ordinary tensor. Such non-centrosymmetric crystal point groups without left-handed symmetries are $\{O,T,C_1,C_n,D_n\}$ with $n={2,3,4,6}$.

However under left-handed symmetries ($\det S=-1$) the transformations of $D^{+}$ differ from those of an ordinary tensor. Equation~\eqref{pseudoT} implies  that under a left-handed symmetry operation the spectrum goes to minus itself: $\{\delta_1,\delta_2,\delta_3\}\rightarrow\{-\delta_1,-\delta_2,-\delta_3\}$. Therefore the only way that it remains invariant as a set, is if it has the form: $\{\delta_1,\delta_2,\delta_3\}=\{\delta,0,-\delta\}$. In such case the eigenvectors would be forced to transform as $S{\bf e}_1=\pm{\bf e}_3$, $S{\bf e}_3=\pm{\bf e}_1$, $S{\bf e}_2=\pm{\bf e}_2$. Therefore, any crystal with a left-handed symmetry and an n-fold rotation axis with $n\geq3$ will force the tensor $D^{+}$ to identically vanish, since such n-fold rotation would additionally force the eigenvectors contained within the invariant plane to be degenerate. For a mirror symmetry, the null eigenvector has to be parallel to the mirror plane, and the eigenvectors with opposite eigenvalues must be at $\pi/4$ angles from the mirror plane, so that they are swapped under the mirror operation. Therefore, the only non-centrosymmetric crystals with mirror symmetries that allow for a non-zero $D^{+}$ can be seen to be $C_{1v}$ and $C_{2v}$~\footnote{Although $C_{2h}$ has only two-fold rotations, the fact that the rotation axis is perpendicular to the mirror plane forces all the elements to vanish.}. For $C_{1v}$ symmetry $D^{+}$ has two independent parameters which can be taken to be the positive eigenvalue and the orientation of the null eigen-vector within the mirror plane. For $C_{2v}$ there is only one independent parameter, which can be taken to be the positive eigenvalue, since the null eigenvector is forced to lie along the rotation axis. In addition the crystal point group $S_4$, which contains a single left-handed four-fold roto-reflection symmetry, allows for a non-zero $D^{+}$, whose null eigenvector is forced to lie along the rotoreflection axis. $D^{+}$ has two independet parameters for $S_4$, which can be taken to be the positive eigenvalue and the orientation of the corresponding eigenvector within the roto-reflection plane.  

\noindent
{\color{blue}{\em Berry curvature dipole in two dimensions}}---In a two-dimensional crystal the Berry curvature behaves as a pseudoscalar (only the component perpendicular to the plane is non-zero), hence the Berry curvature dipole behaves as a {\it pseudo-vector} contained in the two-dimensional plane:
\be\label{Da}
D_{a}=\int_k   f_0 \ (\partial_a \Omega_z).
\ee
This vector has units of length. Therefore symmetry constraints are more severe in two-dimensions. In fact, the largest symmetry of a 2D crystal that allows for a non-vanishing Berry curvature dipole is a single mirror line (i.e. a mirror plane that is orthogonal to the 2D crystal), which would force $D_{a}$ to be orthogonal to it. In vector notation the current can be written as:
\begin{eqnarray}\label{D2D}
\vec{j}^0&=&\frac{e^3\tau}{2 (1+i\omega\tau)} \hat{z}\times\vec{\mathcal{E}}^*(\vec{D}\cdot\vec{\mathcal{E}}), \nonumber \\
\vec{j}^{2\omega}&=&\frac{e^3\tau}{2 (1+i\omega\tau)} \hat{z}\times\vec{\mathcal{E}}(\vec{D}\cdot\vec{\mathcal{E}}). 
\end{eqnarray}
The presence of a single mirror symmetry would force the linear conductivity tensor to have its principal axes aligned with the mirror line. Consequently, according to Eq.\eqref{D2D}, when the driving electric field is aligned with the direction of the Berry curvature dipole vector, $\vec{D}$, all the current that flows orthogonal to it would arise solely from the Berry curvature dipole term.

%On the other hand we can do the analysis by decomposing the 3D tensor version from the previuos section. One finds that since only the $D_{xz}$, $D_{yz}$, components are non-zero, the antysymmetric part is non-zero if and only if the symmetric part is non-zero. Therefore the allowed crystals must be in the intersection of the two sets of symmetry groups so that it allows simultaneously for the symmetric and antisymmetric part to be non-zero. These are $\{C_n,C_{1v},C_{2v}\}$. $C_{2v}$ and $C_2$ could be reconciled with the present analysis if we think that the rotation axis is along the plane containing the 2D crystal. But $\{C_n\}$ with $n\geq3$ can be reconciled with the present analysis. This is probably telling us that the analysis of the previous sections has not distilled the final necessary and sufficient symmetries, but is stuck along the way from necessary onto sufficient. 

\noindent
{\color{blue}{\em Candidate materials}}---Berry curvature is often concentrated in small regions in momentum space where two or more bands cross or nearly cross. 
Therefore, Dirac and Weyl materials are excellent candidates to observe the quantum nonlinear Hall effect predicted in this work. Moreover, since this effect requires the dipole moment of Berry curvature, it is advantageous to choose low-symmetry crystals with {\it tilted} Dirac or Weyl point (see below). We propose three classes of candidate materials: topological crystalline insulators, two-dimensional  transition metal dichalcogenides, and three-dimensional Weyl semimetals.      
  
\begin{figure}
\begin{center}
\includegraphics[width=3in]{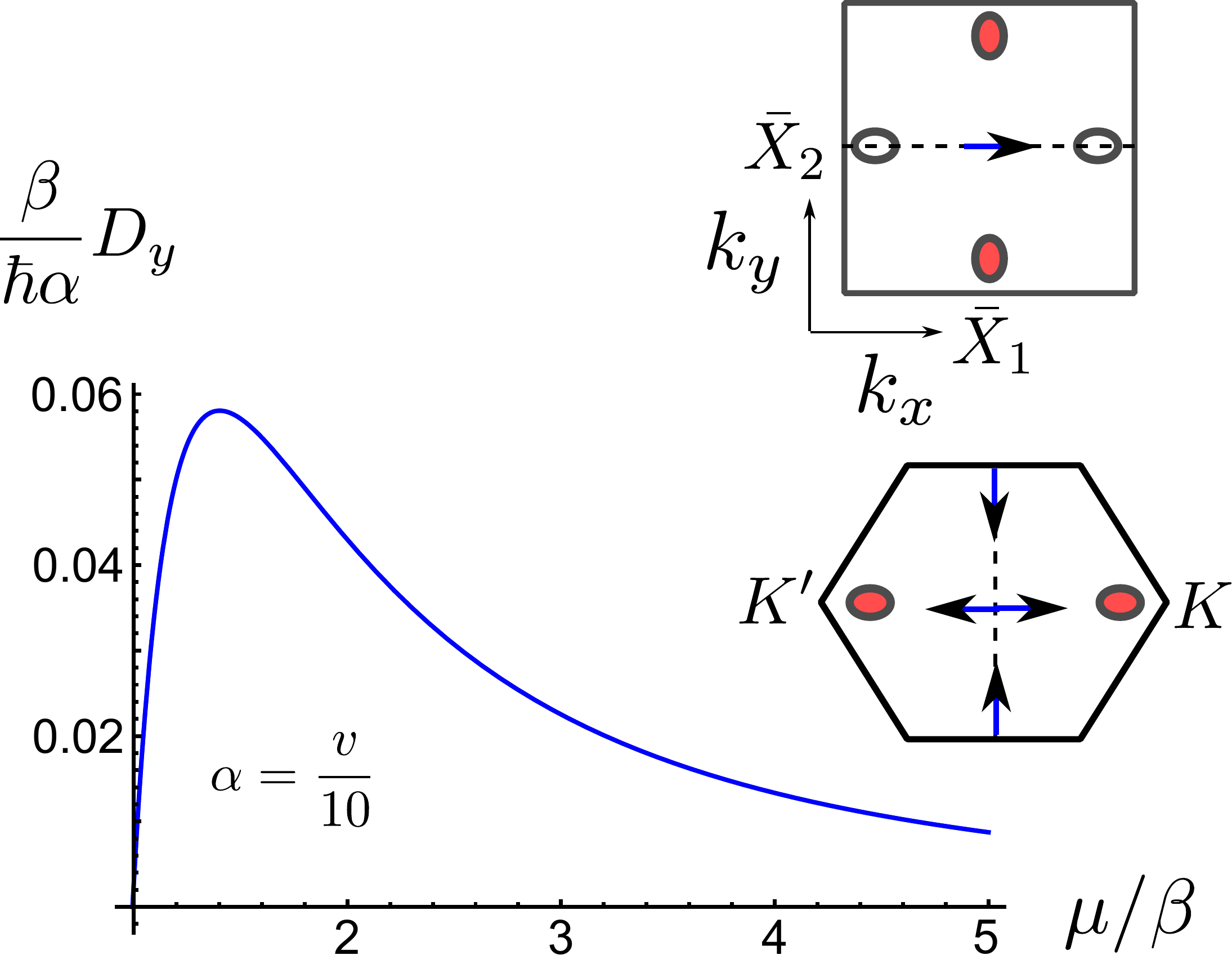}
\caption{(color online) Berry curvature dipole dependence on chemical potential $\mu$. The upper inset  is the surface Brillouin zone of the TCI SnTe or (Pb,Sn)Se. The blue arrow indicates the direction of the ferro-electric distortion. The lower inset illustrates the Brillouin zone of monolayer TMDC with Dirac points shifted away from $K$ and $K'$ by shear strain along the directions indicated by the blue arrows. Red circles and dashed lines indicate massive Dirac points and the surviving mirror symmetry respectively in both insets.}
\label{chi_TCI}
\end{center}
\end{figure}

The surface of topological crystalline insulator (TCI) materials hosts massless Dirac fermions protected by mirror symmetries~\cite{HsiehFu, AndoFu}. In particular, the [001] surface 
of TCIs SnTe, Pb$_{1-x}$Sn$_x$Te and Pb$_{1-x}$Sn$_x$Se, hosts four massless Dirac fermions~\cite{Liu2014} protected by {\it two} mirror symmetries. Pairs of Dirac cones with spin-momentum locking are located near the $\bar{X}$ points of the surface Brillouin zone, forming a Kramers pair. At low temperatures the surface undergoes a structural transition into a ferroelectric  state and one of the mirror symmetries is spontaneously broken~\cite{Okada2013, ARPES}, while the other remains intact.  As a result two of the surface Dirac cones become massive, while the other two remain massless~\cite{Serbyn2014}~(see inset in Fig.~\ref{chi_TCI}). Since the remaining massless Dirac points have vanishing Berry curvature, it is sufficient to consider only the contribution to the Hall current from the two Dirac points that become massive in the distorted crystal structure. They acquire Berry curvatures of opposite signs, because they are mapped into one another by time reversal symmetry. %At small densities away from the Dirac point, so that the Dirac model is reliable, 
The low energy Hamiltonian for the massive Dirac point located at momenta $\pm\Lambda$ away from $\bar{X}_1$ is given by:
\be\label{Hamil}
H_{s\Lambda}=v_x k_x \sigma_y-s v_y k_y \sigma_x+s \alpha k_y+\beta \sigma_z.
\ee
where $s=\pm1$. $\beta$ is the size of the gap opened by the ferroelectric distortion. This low energy theory coincides with that previously considered in the literature~\cite{Liu2014,Okada2013,Serbyn2014}, except for the term proportional to $\alpha$ which produces a finite tilt in the Dirac cones. This tilting effect, which is allowed by symmetry and has been observed in ARPES studies~\cite{Tanaka2012}, is required for a nonzero Berry curvature dipole (see below). 

The dispersion relation for the titled massive Dirac electron is found to be: 
\be
\varepsilon_{s}(k)=s \alpha k_y+{\rm sign}(\mu) (\beta^2+v_x^2 k_x^2+v_y^2 k_y^2)^{1/2}, 
\ee 
where $\mu>0$ ($\mu<0$) for conduction (valence) band. The Berry curvature can be found, from Eq.~\eqref{Berry}, to be:
\be
\Omega_{s}=\frac{{\rm sign}(\mu)}{2}\frac{s v_x v_y \beta}{(\beta^2+v_x^2 k_x^2+v_y^2 k_y^2)^{3/2}}.
\ee
At zero temperature the Berry curvature dipole is:
\be\label{Dtci}
D_{a}= \int_{\varepsilon_{s}(k)<\mu}   \partial_b \Omega.
\ee
%\be\label{chi2D}
%\chi_{abc}=\varepsilon_{ac}  \frac{e^3}{\hbar^2} \frac{\tau}{1+i \omega \tau} \int_{\varepsilon_{s}(k)<\mu}   \partial_b \Omega.
%\ee
%\begin{equation}\label{xx}
%\begin{split}
%&\frac{k_x'^2}{\gamma_x^2}+\frac{(k_y'+s k_0)^2}{\gamma_y^2}=1,\\
%\gamma_x=\frac{\gamma}{v}, \ &\gamma_y=\frac{\gamma}{\sqrt{v^2-\alpha'^2}},\ k_0=\frac{\mu \alpha'}{v^2-\alpha'^2},\\
%v=\sqrt{v_x v_y}&, \ \gamma=\sqrt{\mu^2+(v^2-\alpha'^2)k_0^2-\beta^2},\  \alpha'=\sqrt{\frac{v_x}{v_y}} \alpha.\\
%\end{split}
%\end{equation}
%\begin{equation}
%,
%\end{equation}
This integral can be computed by performing first an area-preserving-transformation: $k_y'=\sqrt{v_y/v_x} k_y$, $k_x'=\sqrt{v_x/v_y} k_x$, and by noting that the Fermi surface is an ellipse in the primed coordinates, namely, that the equation $\epsilon_s(k)=\mu$ implies that: $k_x'^2/\gamma_x^2+(k_y'+s k_0)^2/\gamma_y^2=1$. Where $\gamma_x=\gamma/v$,  $\gamma_y=\gamma/\sqrt{v^2-\alpha'^2}$, $k_0=\mu \alpha'/(v^2-\alpha'^2)$,
 $v=\sqrt{v_x v_y}$, $\gamma=\sqrt{\mu^2+(v^2-\alpha'^2)k_0^2-\beta^2}$,  $\alpha'=\alpha \sqrt{v_x/v_y}$. The condition $v^2>\alpha'^2$ is equivalent to $v_y^2>\alpha^2$ and is needed for the stability of the Dirac cones in the presence of the tilt term. The condition $\mu^2+(v^2-\alpha'^2)k_0^2>\beta^2$ states that the chemical potential is outside the gap, so that there is a finite density of massive Dirac fermions. 

The surviving mirror symmetry, that takes $k_y\rightarrow -k_y$, dictates that only the y-component of the Berry curvature dipole is non-zero, and is found to be:
\begin{equation}\label{Dy}
D_y=\frac{3 v^2 n \beta \alpha |\mu|(1+u^2) }{\bigl[\mu^2 (1+u^2)(1+2 u^2)-u^2 \beta^2\bigr]^{5/2}},
\end{equation}
%\be\label{chi2D}
%\chi_{abc}=\delta_{b,z} \varepsilon_{ac}  \frac{e^3}{\hbar^2} \frac{\tau }{1+i \omega \tau} \frac{3 n \beta \alpha |\mu|(1+\frac{\alpha'^2}{v^2-\alpha'^2}) }{\Bigl[\mu^2 \bigl(1+\frac{\alpha'^2}{v^2-\alpha'^2}\bigr)\bigl(1+ \frac{2 \alpha'^2}{v^2-\alpha'^2}\bigr)-\frac{\alpha'^2 \beta^2}{v^2-\alpha'^2}\Bigr]^{5/2}},
%\ee
\noindent where $u=\alpha'/\sqrt{v^2-\alpha'^2}$ and $n=\int_{|\varepsilon_{s}(k)|<|\mu|}d^2k/(2\pi)^2=\gamma_x \gamma_y/4 \pi$ is the absolute value of the carrier density in each of the massive Dirac cones~\footnote{$n$ is not the full carrier density since there will also be a contribution from the massless Dirac cones.}. Each of the massive Dirac cones has identical contributions to $D_y$, giving rise to a factor of 2 already included in Eq.~\eqref{Dy}. This dipole is orthogonal to the ferroelectric displacement direction in the convention we have chosen. The Berry curvature dipole has the same sign for electrons and holes in this system and vanishes when the chemical potential is in the gap of the massive Dirac fermions. The typical scale of $D_y$ for SnTe TCI is $\hbar \alpha/\beta \sim 3$nm, where we used a Fermi velocity of $v_x\approx v_y\approx 4\times10^5$m/s~\cite{Jiang2012}, $\beta\approx 10$meV and $\alpha=0.1 v_x$. The behavior of $D_y$ is depicted in Fig.~\ref{chi_TCI}. 

%%%% FIG 1 %%%%%
%%%%%%%%%%%%

%To close this section we wish to remark that the third-rank tensor $\sum_s \int_k \partial_{abc} \epsilon_s$ which controls the conventional non-linear response (second term in Eq.~\eqref{ja0}) vanishes due to the time-reversal and the surviving mirror symmetry in SnTe TCI gapped dirac cones. Therefore, the only contribution expected is that arising from the Berry curvature dipole. 

Another candidate 2D materials to observe the quantum non-linear Hall effect are monolayer transition-metal dichalcogenides (TMDC). Their large spin-orbit-coupling and lack of an inversion center produces substantial local Berry curvatures~\cite{Xiao2013,Xu2014}. The $C_{3v}$ symmetry of these crystals would force the Berry curvature dipole to vanish. However, uniaxial strain can reduce this symmetry so that a single mirror operation survives, in which case the effect is allowed. In fact, two copies of the model of Eq.~\eqref{Hamil}, each with a different gap, can describe the states near charge neutrality within a $k\cdot p$ model~\cite{Xiao2013}, and when the shear strain is applied along high-symmetry lines (see inset of Fig.~\ref{chi_TCI}). $s=\pm$ would label valleys K and K' in this case. The anisotropic velocity term parametrized by $\alpha$ would be proportional to the shear strain, much in the same way as in strained graphene~\cite{Linnik2012}. For TMDCs one obtains a scale $\hbar \alpha/\beta \sim 0.2$\AA, using a Fermi velocity of $v\approx 4.5\times10^5$m/s, a gap $\beta\approx 1.5$eV and $\alpha=0.1v$.

%\noindent
%{\color{blue}{\em Concluding remarks}}---
Last but not least, the Berry curvature dipole induced non-linear Hall effect should be present in a very large class of three-dimensional non-centrosymetric crystals. An interesting candidate are the recently discovered Weyl semimetals in the TaAs material class~\cite{xi,hasan-th,hasan-expt, ding}. The crystal structure of these materials is non-centrosymmtric 
and has a polar axis, which allows the quantum nonlinear Hall effect as described by Eq.~\eqref{vecj}. When tilted,  
a Weyl point generates a singular configuration of Berry curvature, with a finite dipole moment whose magnitude can be easily estimated from band structure calculations.  
In addition, other polar materials such as BiTeI with a strong Rashba-type spin-orbit coupling~\cite{Nagaosa2012} may also have large Berry curvature dipole moments. 
These three-dimensional Weyl and Rashba materials provide promising platforms for the observation of the quantum nonlinear Hall effect.

\noindent
{\color{blue}{\em Acknowledgements.}}---
IS is supported by the Pappalardo Fellowship. LF is supported by the DOE Office of Basic Energy Sciences, Division of Materials Sciences and Engineering under award DE-SC0010526.

\bibliographystyle{apsrev4-1}

\begin{thebibliography}{100}

\bibitem{LandauLifshitz} L. D. Landau, E. M. Lifshitz, and L. P. Pitaevskii, Statistical Physics (Course of Theoretical Physics, Volume 5), 3rd. Edition (1999).

 %L. Onsager, Phys. Rev. 37, 405 (1931); Phys. Rev. 38, 2265(1931).

\bibitem{Sundaram1999} G. Sundaram and Q. Niu, Phys. Rev. B {\bf 59}, 14915 (1999). For a review see D. Xiao, M-C. Chang, and Q. Niu, Rev. Mod. Phys. {\bf 82}, 1959 (2010), and references therein.

\bibitem{Moore2010} J. E. Moore and J. Orenstein, Phys. Rev. Lett. {\bf 105}, 026805 (2010).

\bibitem{Mahan} G. D. Mahan, Many-Particle Physics, 3rd. Edition (2000).

\bibitem{Gao2014} Y. Gao, S. A. Yang, and Q. Niu, Phys. Rev. Lett. {\bf 112}, 166601 (2014).

\bibitem{Olbrich2014} P. Olbrich, L. E. Golub, T. Herrmann, S. N. Danilov, H. Plank, V. V. Bel’kov, G. Mussler, Ch. Weyrich, C. M. Schneider, J. Kampmeier, D. Grützmacher, L. Plucinski, M. Eschbach, and S. D. Ganichev, Phys. Rev. Lett. {\bf 113}, 096601 (2014).



\bibitem{Haldane2004} F. D. M. Haldane, Phys. Rev. Lett. {\bf 93}, 206602 (2004).

%\bibitem{Fu2011} L. Fu, Phys. Rev. Lett. {\bf 106}, 106802 (2011).
\bibitem{HsiehFu} 
T. H. Hsieh, H. Lin, J. Liu, W. Duan, A. Bansil and L. Fu, Nat. Commun. {\bf 3}, 982 (2012).

\bibitem{AndoFu}
Y. Ando and L. Fu, Annu. Rev. Condens. Matter Phys. {\bf 6}, 361 (2015).

\bibitem{Liu2014} J. Liu, W. Duan, and L. Fu, Phys. Rev. B {\bf 88}, 241303(R) (2013).

\bibitem{Okada2013} Y. Okada, M. Serbyn, H. Lin, D. Walkup, W. Zhou, C. Dhital, M. Neupane, S. Xu, Y. J. Wang, R. Sankar, F. Chou, A. Bansil, M. Zahid Hasan, S. D. Wilson, L. Fu, V. Madhavan, Science {\bf 341}, 1496 (2013).

\bibitem{ARPES} B. M. Wojek, M. H. Berntsen, V. Jonsson, A. Szczerbakow, P. Dziawa, B. J. Kowalski, T. Story, O. Tjernberg,  arXiv:1505.03414

\bibitem{Serbyn2014} Maksym Serbyn and Liang Fu, Phys. Rev. B {\bf 90}, 035402 (2014).

\bibitem{Tanaka2012} Y. Tanaka, Z. Ren, T. Sato, K. Nakayama, S. Souma, T. Takahashi, K. Segawa and Y. Ando, Nature Physics {\bf 8}, 800 (2012).

\bibitem{Xiao2013} D. Xiao, G-B. Liu, W. Feng, X. Xu, and W. Yao, Phys. Rev. Lett. {\bf 108}, 196802 (2012); G-B. Liu, W-Y. Shan, Y. Yao, W. Yao, and D. Xiao, Phys. Rev. B {\bf 88}, 085433 (2013).

\bibitem{Xu2014} For a recent review see X. Xu, W. Yao, D. Xiao and T. F. Heinz, Nat. Phys. {\bf 10}, 343 (2014).

\bibitem{Linnik2012} T. L. Linnik, J. Phys.: Condens. Matter {\bf 24}, 205302 (2012).

\bibitem{xi} H. Weng, C. Fang, Z. Fang, B. A. Bernevig, and X. Dai, Phys. Rev. X {\bf 5}, 011029 (2015).

\bibitem{hasan-th} S-M. Huang,	S-Y. Xu, I. Belopolski, C-C. Lee, G. Chang, B. Wang, N. Alidoust, G. Bian, M. Neupane, C. Zhang, S. Jia,	A. Bansil, H. Lin and M. Z. Hasan, Nat. Comm. {\bf 6}, 7373 (2015). 

\bibitem{hasan-expt} S-Y. Xu, I. Belopolski, N. Alidoust, M. Neupane, G. Bian, C. Zhang, R. Sankar, G. Chang, Z. Yuan, C-C. Lee, S-M. Huang, H. Zheng, J. Ma, D. S. Sanchez, B. Wang, A. Bansil, F. Chou, P. P. Shibayev, H. Lin, S. Jia, and M. Z. Hasan, arXiv:1502.03807. 


\bibitem{ding} B. Q. Lv, N. Xu, H. M. Weng, J. Z. Ma, P. Richard, X. C. Huang, L. X. Zhao, G. F. Chen, C. Matt, F. Bisti, V. Strokov, J. Mesot, Z. Fang, X. Dai, T. Qian, M. Shi, and H. Ding, arXiv:1503.09188.


\bibitem{Nagaosa2012} M. S. Bahramy, B-J. Yang, R. Arita, and N. Nagaosa, Nat. Comms. {\bf 3}, 679 (2012).

\bibitem{Jiang2012} Y. Jiang, Y. Wang, M. Chen, Z. Li, C. Song, K. He, L. Wang, X. Chen, X. Ma, and Q-K. Xue, PRL {\bf 108}, 016401 (2012).
\end{thebibliography}

\end{document}